\documentclass[sigconf,nonacm]{acmart}
\AtBeginDocument{%
  }

\usepackage{amsmath}
\usepackage{amsfonts}
\usepackage{subcaption}
\usepackage{siunitx}
\usepackage{booktabs}
\usepackage{acronym}
\usepackage{cleveref}
\usepackage{nicematrix}
\usepackage{circledsteps}
\usepackage[dvipsnames]{xcolor}
\usepackage[most]{tcolorbox}

\DeclareMathOperator*{\minimize}{minimize}
\DeclareMathOperator*{\subjectto}{subject\ to}

\newcommand{\Dplan}{D_{\mathrm{plan}}}    
\newcommand{\Dcert}{D_{\mathrm{cert}}}    
\newcommand{\Ddepth}{D_{\mathrm{depth}}}  

\newcommand{\softbox}[1]{\tcbox[colback=black!10!white, colframe=black!25!white, boxrule=0.5pt, arc=0.125mm, top=0.125pt, bottom=0.125pt, left=0.125pt, right=0.125pt, on line]{#1}}
\newcommand{\stM}{\softbox{\textcolor{YellowOrange}{\texttt{\textbf{M}}}}}
\newcommand{\stA}{\softbox{\textcolor{ForestGreen}{\texttt{\textbf{A}}}}}
\newcommand{\stT}{\softbox{\textcolor{blue}{\texttt{\textbf{T}}}}}
\newcommand{\stR}{\softbox{\textcolor{red}{\texttt{\textbf{R}}}}}

\begin{document}

\title{Energy Aware Optimization of Non-Deterministic Task-Based Intermittent Computing Nodes with \texttt{Thimble}}

\author{John Madden}
\authornote{Both authors contributed equally to this research.}
\email{jtmadden@ucsc.edu}
\orcid{{0000-0002-7926-4951}}
\correspondingauthor
\affiliation{%
  \institution{University of California, Santa Cruz}
  \city{Santa Cruz}
  \state{California}
  \country{USA}
}

\author{Morgan Masters}
\authornotemark[1]
\email{mwmaster@ucsc.edu}
\orcid{0009-0004-2788-8051}
\correspondingauthor
\affiliation{%
  \institution{University of California, Santa Cruz}
  \city{Santa Cruz}
  \state{California}
  \country{USA}
}

\author{Alec Levy}
\email{alevy1@ucsc.edu}
\affiliation{%
  \institution{University of California, Santa Cruz}
  \city{Santa Cruz}
  \state{California}
  \country{USA}
}

\author{Colleen Josephson}
\email{cjosephson@ucsc.edu}
\affiliation{%
  \institution{University of California, Santa Cruz}
  \city{Santa Cruz}
  \state{California}
  \country{USA}
}

\author{Steve McGuire}
\email{stmcguir@ucsc.edu}
\affiliation{%
  \institution{University of California, Santa Cruz}
  \city{Santa Cruz}
  \state{California}
  \country{USA}
}

\renewcommand{\shortauthors}{Madden and Masters et al.}

\begin{abstract}

Intermittent computing nodes run on harvested energy, which they store in capacitors and spend in short bursts of computation.
Choosing those capacitors is a trade-off: large capacitors can power large tasks but leak more current, wasting harvested energy, while small capacitors leak less but cannot power large tasks.
Recent work, LEACS, showed that selecting capacitors and mapping tasks to them at compile time improves inference throughput.
However, LEACS optimizes the mapping for a predetermined workload, whereas sensor nodes often run data-dependent workloads in which the next task depends on the outcome of earlier ones.
We present \texttt{Thimble}, an energy storage system that maps a node's tasks to capacitors at runtime.
\texttt{Thimble} models the workload as a Markov chain over tasks and assigns each task to a capacitor by weighing capacitor leakage, the task's energy demand and, when it replans, the likelihood of each future task.
We compare \texttt{Thimble} against a reimplementation of LEACS, which ranks every static plan by simulation.
When the training trace resembles deployment, \texttt{Thimble}'s single design-time solve comes within 0.3--4\% of LEACS's throughput.
When the deployment conditions drift from the characteristics of the training set, LEACS's throughput-focused static plans can fail sharply.
On the lowest-power source we consider, LEACS's plan stalls on deployment traces that harvest less than its training trace, and on some never wakes; \texttt{Thimble} avoids this, with 22--29\% higher throughput at the reference workload.
A sensitivity sweep over 25 workload profiles on that source shows that \texttt{Thimble} with depth replanning leads LEACS by 22--42\% throughout, and what run-time remapping adds: replanning at a fixed interval matches or improves on the static map at 21 of 25 points, by up to 7\%.
\end{abstract}


\begin{teaserfigure}
    \centering
    \includegraphics[width=\linewidth]{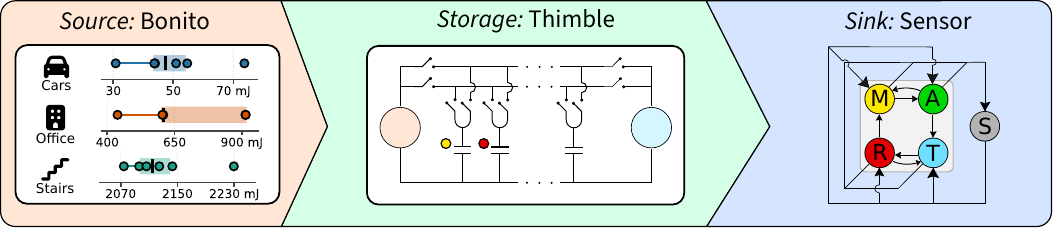}
    \caption{
        Overview of our energy storage system for intermittent computing systems to minimize the amount of energy lost due to capacitor leakages.
        Energy harvested is stored in assigned capacitors based on upcoming tasks and can adjust assigned capacitors based on changing conditions.
        Power is consumed by a model intermittent computing node with stochastic transitions between the states: Measure (\stM{}), Analyze (\stA{}), Transmit (\stT{}), Receive (\stR{}), and Sleep.
    }
    \Description{Visual abstract of a power harvesting intermittent sensing platform. It features three sections. From left to right, there is a representation of harvested power feeds. These feed into a circuit diagram which catches power in capacitors and discharges them based on an task-supply assignment and reassignment method which is presented as the novelty of the paper.}
    \label{fig:teaser}
\end{teaserfigure}

\maketitle

\section{Introduction}

Wireless sensor networks (WSNs) promise to move sensing and analysis onto tiny devices across homes, infrastructure, and outdoor environments.
Maintaining power supplies in the nodes of a WSN has been their weakness in practice: wiring is expensive and easy to damage, and batteries must be replaced or left in the field as e-waste.

Batteryless intermittent platforms typically store harvested energy in capacitors: they charge until a capacitor reaches a turn-on voltage, compute until it falls to a turn-off voltage, and shut down to recharge~\cite{ahmed2024internet,lucia2017snapl}.
Capacitor choice then shapes performance.
A large capacitor finishes an expensive task in one active cycle but recharges slowly and leaks more; a small one recharges quickly and leaks little but cannot power large tasks~\cite{hester2015ufop,barjami2026leacs}.
Multi-capacitor front ends switch among several buffers~\cite{hester2015ufop,colin2018capybara,yang2021morphy}, which raises two questions: which capacitors should a design carry, and which should power each task?

Barjami \emph{et al.}\ showed how much the answer matters~\cite{barjami2026leacs}.
Their tool, LEACS, selects at most $M$ capacitors from a catalog and maps each execution unit of neural network inference to one of them at compile time, raising throughput by $1.59\times$ on average and up to $3.4\times$.
The fixed map assumes a fixed, data-independent sequence of execution units, which most sensor nodes do not have.
A camera node that counts parked bicycles and transmits only when the count changes~\cite{stenkamp_short_2026} spends energy at random, because what it runs next depends on what it saw.
We model such a node as a Markov chain over tasks (\Cref{fig:bh-plane}).
Execution is then a walk into a tree of possible task sequences that the data chooses.

We present \texttt{Thimble}, which selects a capacitor bank at design time, as LEACS does, and re-solves the task-to-capacitor map at run time.
Tasks are atomic and restart after a power failure~\cite{lucia2015dino,maeng2017alpaca}, so we reason in energy rather than time.
Three choices keep the problem tractable and its guarantees sound.
First, near-term tasks are certified against stored energy with no harvest assumed and leakage bounded above, so a near-term assignment that completes in the model completes in situ; later tasks need only fit within a full charge.
Second, within the optimization program, every certificate is a linear constraint and each plan is one binary program with a convex quadratic objective.
Third, the spread of an exact energy forecast over the task tree sets when to replan.
We compare \texttt{Thimble} with and without depth replanning against LEACS trained on the same trace.
We contribute:
\begin{itemize}
    \item An exact forecast of the energy a node draws over its task tree (\S\ref{sec:forecast}), which sets the replanning interval (\S\ref{sec:replan}) and the visits a plan allots each task (\S\ref{sec:problem}).
    \item A one-shot assignment program for design-time selection and run-time remapping, with a two-part energy certificate and a board fixed at design time (\S\ref{sec:problem}).
    \item A SPICE testbed~\cite{ngspice,pyspice} with leakage, switching, and explicit harvest-routing, driven by recorded harvesting traces~\cite{geissdoerfer2022bonito}, and a replication of the LEACS baseline (\S\ref{sec:design}).
    \item An evaluation using starved and well-powered sources showing when storage policy matters, and that simulation-ranked capacitor selection is brittle when deployment harvest differs from training (\S\ref{sec:results}).
\end{itemize}

\section{Background}

\paragraph{Energy harvesting}
Harvesting opens unique deployment scenarios for WSN nodes, such as: soil sensors powered by thermal gradients~\cite{ikeda_soil-monitoring_2020} or microbial fuel cells~\cite{yen_soil-powered_2024}, bridges instrumented with piezoelectric harvesters~\cite{mattauch_design_2025}, and railway tags powered by RF from passing trains~\cite{li2021rf}.
Harvested power is erratic in source-specific ways: solar follows the day and the weather~\cite{naderi2025solar}, piezoelectric harvesting yields short bursts, and floor-embedded office solar yields the most but goes dark for minutes when occluded~\cite{geissdoerfer2022bonito}.
The best storage therefore depends on the source: LEACS's best single capacitor was \SI{19}{\micro\farad} on the jogging trace but \SI{78}{\micro\farad} in the office~\cite{barjami2026leacs}.

\paragraph{Energy storage in intermittent systems}
Task-based systems split programs into atomic tasks whose effects commit only on completion~\cite{lucia2015dino,maeng2017alpaca}; a task whose capacitor cannot hold enough energy will brownout and lose its progress, which CleanCut checks at compile time~\cite{colin2018cleancut}.
\Cref{tab:systems} summarizes how systems decide what storage to use.

\begin{table}[t]
  \centering
  \footnotesize
  \caption{How intermittent systems decide energy storage.}
  \label{tab:systems}
  \begin{tabular}{@{}l >{\raggedright\arraybackslash}p{5.4cm} l@{}}
    \toprule
    System & Storage decision & When \\
    \midrule
    Single buffer & one capacitor, sized by hand & design \\
    UFoP~\cite{hester2015ufop} & per-load capacitors & design \\
    Capybara~\cite{colin2018capybara} & bank per declared energy mode & run time \\
    Stash~\cite{alsubhi2024stash} & adaptive multi-capacitor front end & run time \\
    CleanCut~\cite{colin2018cleancut} & checks termination, given capacitor & compile \\
    LEACS~\cite{barjami2026leacs} & capacitor set and static map, ranked by replay & compile \\
    \textbf{Thimble} & set at design time; map re-solved on forecast overdraw & both \\
    \bottomrule
  \end{tabular}
\end{table}

\paragraph{Adapting the workload}
Other systems adapt the execution of tasks based on available energy.
REHASH and Protean adapt the application to harvesting conditions with runtime heuristics~\cite{bakar2021rehash,bakar2022protean}, and send-on-delta reporting and data compression trade the frequency or fidelity of transmissions for energy~\cite{miskowicz2006sod,martinezsanchez2025sod,saidani2020lossless,correa2022lossy,xiao2017cnmstp}.
In our model (\S\ref{sec:state-machine}), such techniques set how often a node skips a transmission and how often it retries one; \texttt{Thimble} takes these branch probabilities as given and decides how to store and spend energy.



\section{Model Derivations}

\subsection{State Machine Analysis}
\label{sec:state-machine}

\begin{figure}[t]
    \centering
    \includegraphics[width=\linewidth]{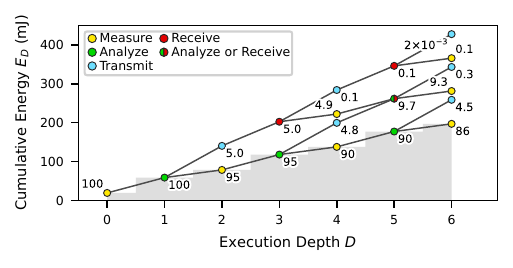}
    \caption{
        Markov chain tree for the branch probabilities $a=0.95$ and $b=0.05$ (\S\ref{sec:state-machine}), showing the evolution of node states with the execution depth $d$, the number of transitions since the last replan.
        As the depth increases, the uncertainty of the amount of energy consumed increases.
        Labels give the probability of each path, in percent.
        This uncertainty sets when \texttt{Thimble} replans (\Cref{fig:triggers}).
    }
  \label{fig:bh-plane}
  \Description{A figure displaying the development of trajectories, described by a Markov chain, on a plane defined by the number of Markov chain iterations applied (x-axis) and the available energy stored in a capacitor array (y-axis).}
\end{figure}

We define a representative sensor node state machine that cycles through the states: Measure \stM{}, Analyze \stA{}, Transmit \stT{}, and Receive \stR{}, under the following rules:\footnote{Colors from the states of~\Cref{fig:teaser} are propagated through the sprites representing state names and their abbreviations, to aid interpretability in this section.}
\begin{enumerate}
    \item \stM{} always advances to \stA{};
    \item \stT{} always advances to \stR{};
    \item \stA{} returns to \stM{} with probability $a\in(0,1)$ or advances to \stT{} with probability $1-a$; and
    \item \stR{} returns to \stT{} with probability $b\in(0,1)$ or advances to \stM{} with probability $1-b$.
\end{enumerate}
For example, consider $a$ as the probability that a send-on-delta reporting scheme rejects a measurement for transmission and $b$ as the probability of a transmission failure.
These probabilities are constant throughout the simulation and can be inferred from datasets or through the experimental setup.
This machine is shown in the rightmost bubble of~\Cref{fig:teaser}, and the transition matrix, $P$, of the equivalent Markov chain is:
\begin{equation}
    \label{eq:mc_4x4}
    P = \;\;
    \begin{NiceMatrix}
        \stM{} &  \stA{} &  \stT{} &  \stR{} &  \\
        0 & 1 & 0 & 0 &  \stM{} \\
        a & 0 & (1-a) & 0 &  \stA{} \\
        0 & 0 & 0 & 1 &  \stT{} \\
        (1-b) & 0 & b & 0 &  \stR{}
    \CodeAfter
        \SubMatrix[{2-1}{5-4}]
\end{NiceMatrix}
\end{equation}
$P$ is row-stochastic, so a distribution $\underline{q}_d$ over the states after $d$ transitions propagates as a row vector, $\underline{q}_{d+1}^T=\underline{q}_d^T P$.
Every transition moves between $\{\stM{},\stT{}\}$ and $\{\stA{},\stR{}\}$, so the chain has period~2.
Since only steps out of \stA{} and \stR{} branch, the number of trajectories doubles every two steps (\Cref{fig:bh-plane}).
Each state runs one task: we write $\mathcal{U}=\{u_1,\dots,u_N\}$ for the tasks, here $N=4$, and index the rows and columns of $P$ by task, so $P_{ii'}$ is the probability that task $u_{i'}$ runs next after task $u_i$.

\subsection{Energy Costs}
\label{sec:costs}

Task $u_i$ draws power $p_i$ for $\tau_i$ seconds, so one run needs $E_i=p_i\tau_i$ joules of energy.
The energy comes from a catalog $\mathcal{C}=\{c_1,\dots,c_K\}$ of $K$ capacitors.
Capacitor $c_j$, with capacitance $C_j$ and cutoff and ceiling voltages $V_{\min,j}$ and $V_{\max,j}$, holds usable energy $S_j$ at its present voltage $V_j$, and $F_j$ when charged to $V_{\max,j}$:
\begin{equation}
  S_j=\tfrac12 C_j\big(V_j^2-V_{\min,j}^2\big)
  \quad\text{and}\quad
  F_j=\tfrac12 C_j\big(V_{\max,j}^2-V_{\min,j}^2\big).
  \label{eq:energy}
\end{equation}
Datasheets model the leakage current of $c_j$ at voltage $V$ as $\max(\kappa C_jV,\;I_0)$, with a leakage coefficient $\kappa$ and a current floor $I_0$, typically $\kappa=\SI{0.01}{\per\second}$ and $I_0=\SI{3}{\micro\ampere}$.
Both \texttt{Thimble} and LEACS~\cite{barjami2026leacs} use this model evaluated at the capacitor's voltage.
Leakage grows with voltage, so over a run of $u_i$ started at voltage $V$, $c_j$ loses at most $l_{ij}(V)=\tau_i V\max(\kappa C_jV,\,I_0)$, and the run costs:
\begin{equation}
  e_{ij}(V)=E_i+l_{ij}(V).
  \label{eq:costs}
\end{equation}
A switching cost $\gamma_j$ for capacitor $c_j$ can be added to~\eqref{eq:costs}; we omit it here.

\subsection{Forecasting Energy on the Task Tree}
\label{sec:forecast}

Let $u_\rho$ be the task running when \texttt{Thimble} plans (the root, $\rho$), and let the execution depth $d$ count the transitions since then.
Propagating the chain recursively gives the exact joint distribution $\pi_d(i,E)$ of the task $u_i$ running at depth $d$ and the energy $E$ consumed from the root through $u_i$:
\begin{equation}
  \pi_0(i,E)=[\,i=\rho\,]\,[\,E=E_\rho\,],\qquad
  \pi_{d+1}(i',E+E_{i'})\mathrel{+}=\pi_d(i,E)\,P_{ii'},
  \label{eq:leaf}
\end{equation}
where $[\cdot]$ is 1 if its argument holds and 0 otherwise, $\pi_{d+1}$ starts at zero, and the update runs over every $(i,E)$ in the support of $\pi_d$ and every successor $u_{i'}$.
The spread of the forecast at depth $d$ is $\sigma_d$, the standard deviation of $E$ under $\pi_d$; \Cref{fig:triggers} shows it growing with depth.

\subsection{Assignment Program}
\label{sec:problem}

At most $M\le K$ capacitors from the catalog go on the board $\mathcal{C}'\subseteq\mathcal{C}$, and a map $\mu:\mathcal{U}\to\mathcal{C}'$ assigns every task to one board capacitor.

The decision vector $\underline{z}=[\underline{x}^T\ \underline{y}^T]^T\in\{0,1\}^{K(N+1)}$ stacks the selection $\underline{x}\in\{0,1\}^K$, with $x_j=1$ if $c_j$ is on the board, and the one-hot assignments $\underline{y}=[\underline{y}_1^T\cdots\underline{y}_N^T]^T$, $\underline{y}_i\in\{0,1\}^K$, with $y_{ij}=1$ if $\mu(u_i)=c_j$.
We write $\mathbf{1}_K$ and $\mathbf{0}_K$ for the all-ones and all-zeros vectors of length $K$, likewise for other lengths, and drop the subscript where the length is clear.

Operators extract the assignments ($\Phi_1$) and the selection ($\Phi_2$, and $X$), replicate the capacitor selection into each assignment block ($\Phi_3$), sum assignments per task ($\Phi_4$), and sum tasks per capacitor ($\Phi_5$). 
For the definitions of these operators, see~\Cref{sec:appendix}.

A plan allots $n_i$ visits to task $u_i$: either one, or the expected number of visits over the next $\Dplan$ transitions from the root, the planning horizon, rounded up,
\begin{equation}
  n_i=\Big\lceil\textstyle\sum_{d=0}^{\Dplan}(\underline{\delta}_\rho^TP^d)_i\Big\rceil,
  \label{eq:visits}
\end{equation}
where $\underline{\delta}_\rho\in\{0,1\}^N$ indicates the root task, with entries $[\,i=\rho\,]$.
The weights $w_{ij}=n_i\,e_{ij}(V_j)$, the energy that the allotted visits to $u_i$ draw from $c_j$, form $\underline{w}=[\mathbf{0}_K^T\;\underline{w}_1^T\cdots\underline{w}_N^T]^T$, laid out like $\underline{z}$.

\paragraph{Certificate}
Let $\Omega$ be the set of task sequences of $\Dcert$ transitions from the root, the certificate horizon, and $\mathcal{H}\subseteq\mathcal{U}$ the set of tasks they contain.
Tasks in $\mathcal{H}$ must run from stored energy and later tasks must fit in a full charge, so the feasibility mask is
\begin{equation}
  f_{ij}=\begin{cases}
    \big[\,e_{ij}(V_j)\le S_j\,\big] & u_i\in\mathcal{H},\\
    \big[\,e_{ij}(V_{\max,j})\le F_j\,\big] & \text{otherwise},
  \end{cases}
  \label{eq:f}
\end{equation}
with $\underline{f}$ laid out like $\underline{z}$ and its first $K$ entries set to one.
Because a trajectory may visit one capacitor more than once, each $\omega\in\Omega$ must also fit stored energy jointly, with $m_\omega(i)$ the number of times $\omega$ visits $u_i$.
LEACS's feasibility check is the ``otherwise'' branch of~\eqref{eq:f}, exclusively.

\paragraph{Design time and run time}
At design time, \eqref{eq:program} runs once with every capacitor at full charge, $V_j=V_{\max,j}$, so that every task is certified against a full charge, with one visit per task, $n_i=1$, and without the last line; the capacitors its map uses become the board $\underline{x}^0\in\{0,1\}^K$.
The chain thus enters only at run time.
At run time, $V_j$ is each capacitor's present voltage, replans allot visits by~\eqref{eq:visits}, and $X\underline{z}=\underline{x}^0$ freezes the hardware, so replans change only the map.

\paragraph{Program}
Given the board limit $M$ and the weight matrix $W=\Phi_5\,\mathrm{diag}(\underline{w})/\max_{ij}w_{ij}$, \texttt{Thimble} solves
\begin{equation}
  \begin{aligned}
    \minimize_{\underline{z}}\;\; & \lVert W\underline{z}\rVert_2^2\\
    \subjectto\;\; & \Phi_4\Phi_1\underline{z}=\mathbf{1}_N,\quad (\Phi_1-\Phi_3)\underline{z}\le\mathbf{0},\\
    & \Phi_1\underline{z}\le\underline{f},\quad \mathbf{1}^T\Phi_2\underline{z}\le M,\\
    & \textstyle\sum_i m_\omega(i)\,e_{ij}(V_j)\,y_{ij}\le S_j x_j\quad\forall\,\omega\in\Omega,\ j,\\
    & X\underline{z}=\underline{x}^0.
  \end{aligned}
  \label{eq:program}
\end{equation}

The objective balances load via a least-squares objective and breaks the symmetry of an otherwise large equivalence class by weighting the task assignments subvector of the decision vector with the energy costs incurred by every possible task-to-capacitor assignment.

With every task certified against a full charge, the first two lines are LEACS's constraints in operator form; the third line and the stored-energy branch of the mask are \texttt{Thimble}'s additions. 
The first line of constraints maps every task to one selected capacitor.
The second applies the certificate mask and the board capacitor limit.
The third certifies every near-term trajectory against stored energy with no harvest; harvest that does arrive only adds margin.
The last line holds the board fixed after the design-time solve.
$W^TW\succeq0$, so the objective is convex, but the variables are binary; we solve~\eqref{eq:program} with SCIP~\cite{Achterberg2009scip}.

\subsection{Replanning}
\label{sec:replan}

\begin{figure}
    \centering
    \includegraphics[width=\linewidth]{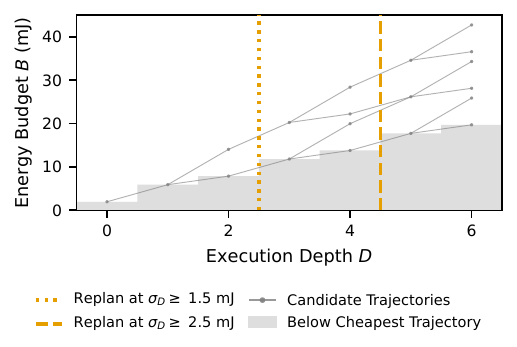}
    \caption{
        Depth replanning (vertical lines) replans at the first depth $d$ where the standard deviation of the forecast energy reaches a threshold, $\sigma_d\geq\sigma_{\max}$; depth replanning sets this depth directly.
        }
    \label{fig:triggers}
    \Description{A figure depicting several depth triggering decision boundaries, overlaid on the earlier tree figure (\Cref{fig:bh-plane}).}
\end{figure}

We consider a triggering scheme which quantifies uncertainty in the forecast energy state.
Represented as vertical, gold lines in~\Cref{fig:triggers}, it fires at the first depth at which the spread $\sigma_d$ of~\S\ref{sec:forecast} reaches a threshold $\sigma_{\max}$.
Because $\sigma_d$ depends only on the root task and the chain, not on the energy stored, this trigger fires a fixed number $\Ddepth$ of transitions after each replan; we set that number directly, replanning every $\Ddepth=2$ transitions, and call this \textbf{depth replanning}.


A replan re-solves~\eqref{eq:program} from the present charge state, rooted at the running task, which gives the new map the whole task to charge the capacitors it assigns to successors.
An infeasible solve reverts to the design-time map.
We compare this trigger with LEACS and with a variant of \texttt{Thimble} that never replans (\Cref{tab:runtime}).

\section{Simulated Experiment Design}
\label{sec:design}

We simulate \texttt{Thimble} in Python with PySpice~\cite{pyspice} and ngspice~\cite{ngspice}.
The simulation can be configured with different capacitors, switching models, and number of power lines.
The power source and sink can be arbitrarily specified, with the timeseries values used to set the power supplied and power consumed at each simulation step.
The source mimics an energy harvester chip, supplying the specified constant power with variable voltage and current.
Similarly, the sink exhibits piecewise constant power draw, as dictated by the state machine, with a cutoff voltage to simulate brownout conditions.
After the simulation completes, timeseries data on power, voltages, and currents are accessible for evaluation; Appendix~\ref{sec:circuit} details the circuit.
We release the simulation framework as an open source tool publicly available on GitHub.

\paragraph{Methods and runtime}
\Cref{tab:runtime} lists the three methods and the rules they share.
LEACS ranks every static plan on a training trace and runs the best one unchanged.
The two \texttt{Thimble} variants share the design-time board and map of~\eqref{eq:program} and differ only in when they replan.

\begin{table}[t]
  \caption{Methods and runtime. Every \texttt{Thimble} variant uses the energy-balancing objective, $M=2$, and the two-part certificate~\eqref{eq:f} over $D_{cert}=1$ transition; replans allot visits by~\eqref{eq:visits} over $\Dplan=2$ transitions, and an infeasible solve reverts to the design-time map.}
  \label{tab:runtime}
  \small
  \begin{tabular}{@{}>{\raggedright\arraybackslash}p{0.25\linewidth}>{\raggedright\arraybackslash}p{0.69\linewidth}@{}}
    \toprule
    \emph{Fixed rules} & \\
    Execution & Each step, a task pauses if its capacitor cannot finish it. \\
    Routing & Harvest goes to the next tasks' capacitors, then the running capacitor, then another board capacitor; the source is on one line at a time. \\
    Wake & A sleeping node wakes when the capacitor of the task it resumes reaches $V_{\max}$, the rule LEACS assumes. \\
    \midrule
    \emph{Method} & \emph{Map} \\
    LEACS & best of 104 static plans feasible under the first two lines of~\eqref{eq:program} at full charge, ranked on the training trace; never replans \\
    \texttt{Thimble}, static & design-time map of~\eqref{eq:program} \\
    \texttt{Thimble}, depth & replans every $\Ddepth=2$ transitions \\
    \bottomrule
  \end{tabular}
\end{table}

\paragraph{Workload}
\Cref{tab:workload} lists the tasks' power draw, measured from a representative WSN platform based on an STM32 microcontroller and transmitting via LoRaWAN, scaled down by a factor of 10$\times$. As the platform did not have an Analyze state, its power draw was represented as twice the cost of \stM{}. The capacitor catalog used one decade of the E6 series values, from \SI{1.0}{} to \SI{6.8}{\milli\farad}.
Based on the selected range, the \SI{1}{\milli\farad} capacitor can run only \stM{}, \SI{1.5}{\milli\farad} also runs \stA{}, and \SI{2.2}{\milli\farad} and up run every task.
At $M=2$ the catalog admits 104 static plans, and at $M=1$ four.

\begin{figure}[t]
    \centering
    \includegraphics[width=0.85\linewidth]{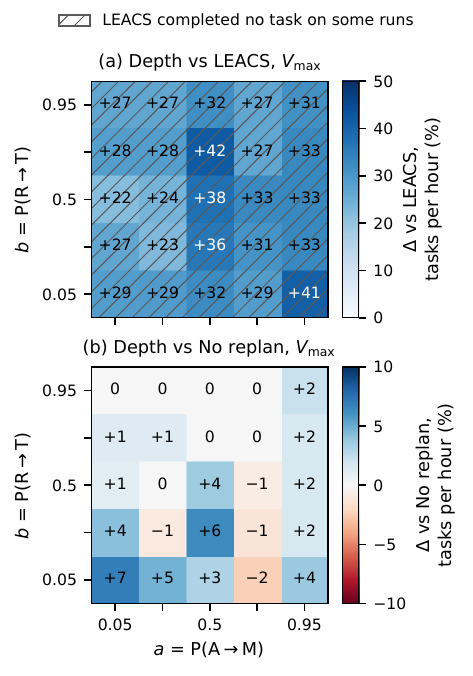}
    \caption{Sensitivity sweep on \texttt{cars}, pooled over held-out traces and seeds.
    (a)~Depth replanning ($\Ddepth=2$) versus LEACS, which keeps the plan it ranked at $a=b=0.5$; hatched cells: LEACS completed no task on some runs.
    (b)~Depth replanning versus \texttt{Thimble}'s static map, which shares its board and, for a given seed, its task sequence.
    All runs charge capacitors to $V_{\max}$ before waking.}
    \Description{Two heatmaps over the (a, b) plane on the cars source. Depth replanning leads LEACS by 22 to 42 percent everywhere, and changes throughput by -2 to +7 percent relative to Thimble's static map.}
    \label{fig:heatmaps}
\end{figure}

The comparison fixes $a=b=0.5$, the center of the plane, where both branches are most random.
There, every stochastic task is equally likely and a node that never stops would complete $3600/0.5585\approx6{,}446$ tasks per hour, its estimated throughput ceiling, where \qty{0.5585}{\s} is the mean task duration of the workload.
We perform a sensitivity sweep over the $(a,b)$-plane to characterize the effect of stochasticity on the compared methods and runtimes (\Cref{fig:heatmaps}).

\begin{table}[htb]
  \caption{Energy requirements for each of the tasks of the \stM\stA\stT\stR{} machine and the storage capacity of the chosen E6 capacitor catalog ($V_{\min}=\SI{1.65}{\volt}$, $V_{\max}=\SI{3.2}{\volt}$).}
  \label{tab:workload}
  \small
  \setlength{\tabcolsep}{4pt}
  \begin{tabular}{@{}rcll c rl@{}}
    \toprule
    \multicolumn{4}{c}{\textbf{Task Profile}} & & \multicolumn{2}{c}{\textbf{Capacitor Catalog}} \\
    \cmidrule(r){1-4} \cmidrule(l){6-7}
    Task & $p_i$ (\si{\milli\watt}) & $\tau_i$ (\si{\second}) & $E_i$ (\si{\milli\joule}) & & $C_j$ (\si{\milli\farad}) & $F_j$ (\si{\milli\joule}) \\
    \midrule
            &       &       &      & & $c_1$ (1.0) & 3.76 \\
    \stM{}  & 3.74  & 0.511 & 1.91 & & $c_2$ (1.5) & 5.64 \\
    \stA{}  & 7.48  & 0.511 & 3.82 & & $c_3$ (2.2) & 8.27 \\
    \stT{}  & 27.69 & 0.285 & 7.89 & & $c_4$ (3.3) & 12.40 \\
    \stR{}  & 6.41  & 0.927 & 5.94 & & $c_5$ (4.7) & 17.67 \\
            &       &       &      & & $c_6$ (6.8) & 25.56 \\
    \bottomrule
  \end{tabular}
\end{table}

\paragraph{Sources}
Harvested power comes from three Bonito sources~\cite{geissdoerfer2022bonito}, which are traces captured from a number of piezoelectric and photovoltaic harvester sources.
They were chosen because they span the regimes that matter: \verb|cars| (six traces), \verb|office| (five) and \verb|stairs| (six), held by zero-order hold when downsampling in time.
Bonito's traces are sized for microfarad storage, so we scale them by 64, the smallest power of two at which, under \texttt{Thimble}, the model platform wakes on every trace.
At this scale, \verb|stairs| nodes run near the throughput ceiling; \verb|jogging|, whose recorded peaks exceed \verb|stairs|' severalfold, would saturate in the same way, and \verb|washer|'s bursty energy profile led to infeasible computational time for simulation, at the same power level as \verb|stairs|, so we omit them.

\paragraph{Protocol and metric}
The first trace of each source trains LEACS, with two random seeds and \SI{1200}{\second} runs; the others are held out, and every method runs on each with three seeds for \SI{3600}{\second}.
The chain draws from one random stream per run, and only when \stA{} or \stR{} exits, so for a given seed every method walks the same sequence of tasks.
Throughput is defined as tasks completed per hour.
We report each \texttt{Thimble} variant's mean throughput over the held-out runs relative to LEACS'.

\paragraph{Sensitivity sweep}
We sweep $a,b\in\{0.05,0.275,0.5,0.725,0.95\}$ on the \verb|cars| source, with two new seeds per held-out trace; \texttt{Thimble}'s model uses the true $(a,b)$ at every point.
LEACS keeps the plan it ranked at $a=b=0.5$, as a compile-time plan would when the workload drifts; re-ranking it at each point would cost 208 runs per point, and in preliminary simulations its best plan barely moved with $(a,b)$.
We report depth replanning's throughput relative to LEACS and to \texttt{Thimble}'s static map, which shares its board and, for a given seed, its task sequence, so the second ratio isolates what run-time remapping adds.

\section{Results}
\label{sec:results} 
\paragraph{Primary experiment.}
\texttt{Thimble}'s program never sees the harvest, so it always chooses the same board, \SI{2.2}{} and \SI{3.3}{\milli\farad}, on every source based on knowledge of task energy requirements and leakages.
The two capacitors are the smallest available which can cover all the tasks with a balanced assignment (i.e. not all on a single capacitor). 
Comparatively, LEACS's ranking chooses boards from \qtyrange{2.2}{6.8}{\milli\farad} to suit each power environment under the metric of maximum throughput over a training dataset.
That helps LEACS where the held-out traces resemble the training trace, but hurts it where they do not.

Table~\ref{tab:results} shows two harvesting regimes. 
On the energy-dense sources, where nodes run close to continuously, \texttt{Thimble} trails LEACS slightly: by 0.3--0.4\% on \texttt{stairs}, where LEACS reaches 93\% of the throughput ceiling, and by 3.4--3.6\% on \texttt{office}, with LEACS throughput at 82\%.
The \verb|cars| training trace is less energy dense and more variable between traces, delivering \SI{1.10}{\milli\watt} against 0.49--\SI{0.72}{\milli\watt} on the held-out traces.
Waiting for the capacitor to charge to $V_{\max}$ before waking, LEACS's \SI{6.8}{\milli\farad} plan completes no tasks on two of the five held-out traces due to high leakage currents preventing capacitors from charging above their threshold.
Alternatively, \texttt{Thimble} completes tasks on all held-out traces, because of the capacitors it selects, and improves throughput over LEACS by \qty{22}{\percent} without replanning and \qty{29}{\percent} with depth replanning.

\begin{table}[t]
  \centering
  \small
  \caption{Throughput at $a=b=0.5$ on held-out traces, in tasks per hour: the mean over every held-out trace and seed (15 runs for \texttt{stairs} and \texttt{cars}, 12 for \texttt{office}). A node that never stops would complete roughly 6{,}446. \emph{Caps.}: LEACS's board; \texttt{Thimble}'s is $c_3, c_4$ on every source. On \texttt{cars}, LEACS completed no task on two of the five traces.}
  \label{tab:results}
  \setlength{\tabcolsep}{4pt}
  \begin{tabular}{@{}l l
                  S[table-format=4.1]
                  S[table-format=4.1]
                  S[table-format=4.1]@{}}
    \toprule
           & \multicolumn{2}{c}{LEACS}
           & \multicolumn{2}{c@{}}{\texttt{Thimble}} \\
    \cmidrule(lr){2-3} \cmidrule(l){4-5}
    Source & Caps. & {Tasks/h} & {Static} & {Depth} \\
    \midrule
    Stairs & $c_3, c_6$ & 6002.5 & 5976.2 & 5983.4 \\
    Office & $c_5, c_6$ & 5261.5 & 5082.3 & 5071.1 \\
    Cars   & $c_6$      &    8.3 &   10.1 &   10.7 \\
    \bottomrule
  \end{tabular}
\end{table}

\paragraph{Sensitivity sweep.}
Across the sweep, LEACS's plan completes no task on at least one held-out seed in every cell, and depth replanning leads it by 22--42\% (\Cref{fig:heatmaps}a).
The sweep uses new seeds, so its center cell need not match \Cref{tab:results}.
Because these ratios are dominated by LEACS stalling, \Cref{fig:heatmaps}b compares depth replanning with \texttt{Thimble}'s static map.

Depth replanning matches or beats the static map in 21 of 25 cells and trails it by at most 2\% in the other four.
Its gains are largest when retransmissions are rare, peaking at 7\% at $a=b=0.05$, and nearly vanish at $b=0.95$ (0--2\%), where retransmissions dominate the workload.

\section{Discussion}
\label{sec:discussion}

\paragraph{One solve against a ranked search}
\texttt{Thimble}'s board capacitor selection and capacitor to task map come from one solve of~\eqref{eq:program}; LEACS's come from many simulations (here: 104 feasible plans $\times$ two seeds).
Where training and deployment harvest agree, the ranked search increases task throughput of LEACS, but the method is brittle to shifts in the deployment energy environment in a way \texttt{Thimble}'s harvest-independent choice was not.

\paragraph{What remapping can do}
Remapping moves tasks between the capacitors on the board, so it can recover energy the current map strands but not storage the board lacks.
On well-powered sources the design-time map rarely lacks the optimal set of capacitors, and remapping has little effect on the performance.
On the energy-scarce \texttt{cars} source, \texttt{Thimble}'s depth replanning shows promise in improving throughput when every Coulomb counts.
The depth replanning trigger was not tuned: the depth interval $\Ddepth$ trades responsiveness against solves.
Sweeping the depth, and testing remapping where the workload drifts from its design point (\Cref{fig:heatmaps}), has the potential to improve performance.

\subsection{Limitations}

\paragraph{A harvest-blind process}
\texttt{Thimble}'s certificate ignores future harvest so that its near-term guarantees hold whatever arrives, and its design-time program likewise sees the tasks but not the harvest, so its board is the same on every source.
Adding harvest statistics (such as a trace's mean intensity or mean burst length, or an energy forecast heuristic) to the design-time program is the natural next step.

Even though a widely accepted circuit simulator was used, there were aspects that do not reflect a real-world implementation that could impact the results. 
At the source, we assume an ideal energy harvester that can supply constant power with no efficiency loss as the voltage and current varies. 
Additionally, real world PFET switches were replaced with a simplified model for simulation speed, so we implicitly assume the cost of switching is negligible compared to other losses.

Although we tried manufacturer-provided capacitor models, they did not model leakage current.
Instead, the simulation implements a simplified leakage model with a constant leakage coefficient $\kappa$, as in~\S\ref{sec:costs} and Appendix~\ref{sec:circuit}.
In reality, capacitors have complex behaviors as $\kappa$ can vary between technologies/manufacturers, and leakage changes over time since last charge \cite{silicon_labs_current_2023}.

\section{Future Work}

\paragraph{Hardware}
Implementing \texttt{Thimble} on embedded hardware would test the simulation against measurement, including its leakage model, and would characterize capacitor leakage under harvesting conditions, which manufacturer models omit.
It would also measure the energy the optimizer itself consumes, to weigh against the leakage it saves.

\paragraph{Richer workloads}
We model a node with four tasks and two probabilistic transitions.
Real nodes have more: a wired bus or a radio stack runs several steps, each of which can fail and retry.
Splitting tasks into smaller atomic units would also let \texttt{Thimble} place each on a smaller, less leaky capacitor.
Beyond a single node, the program could extend federated energy storage~\cite{hester2015ufop} across several harvesters, nodes and peripherals.

\paragraph{Learning the chain}
Our evaluation gives \texttt{Thimble} fixed, known transition probabilities, but in deployment they drift: neighboring nodes change the packet loss behind $b$, and the measured signal changes how often send-on-delta reporting transmits, behind $a$.
Estimating the transition matrix online from observed transitions would let \texttt{Thimble} track these changes.

\section{Conclusion}

\texttt{Thimble} selects a capacitor bank at design time and can remap tasks at run time, from one binary program, guided by an exact forecast of the energy the node draws. 
Without seeing the harvest, its single design-time solve comes within 0.3--4\% of LEACS's simulation-ranked plans on well-powered and saturated sources. 
On the starved \texttt{cars} source, LEACS's training trace is richer than deployment, and its plan stalls on two of five held-out traces. 
\texttt{Thimble} completes tasks on all of them and achieves 22--29\% higher throughput. 
Run-time replanning matters only where storage binds. It changes little on \texttt{stairs} and \texttt{office}. 
On \texttt{cars}, depth replanning adds up to 7\% over the static map, most when retransmissions are rare. 
Adding harvest statistics to the design-time program and tuning the replanning interval are the natural next steps.

\begin{acks}
\end{acks}

%

\bibliographystyle{ACM-Reference-Format}
\bibliography{refs.bib}

\appendix

\section{Operator Definitions}
\label{sec:appendix}

Here $\mathbb{I}_K$ is the $K\times K$ identity and $\mathbf{0}_{K\times N}$ the $K\times N$ zero matrix, likewise for other sizes, $\otimes$ is the Kronecker product, and an unlabeled $\mathbf{0}$ block takes the size that completes its matrix.

$\Phi_1$, the operator which extracts the task-to-capacitor assignments from our primal variable:
\begin{equation}
  \Phi_1=\begin{bmatrix}\mathbf{0}_{K\times K}&\mathbf{0}\\\mathbf{0}&\mathbb{I}_{KN}\end{bmatrix}, \quad \Phi_1 \underline{z} = [\mathbf{0}_K^T \; \underline{y}^T]^T
\end{equation}

$\Phi_2$, the operator which extracts the capacitor selection from our primal variable, and $X$, the first block-row of $\Phi_2$:
\begin{gather*}
  \Phi_2=\begin{bmatrix}\mathbb{I}_{K}&\mathbf{0}\\\mathbf{0}&\mathbf{0}_{KN\times KN}\end{bmatrix}, \quad \Phi_2 \underline{z} = [\underline{x}^T \; \mathbf{0}_{KN}^T]^T, \\
  \\
  X=[\,\mathbb{I}_K\;\;\mathbf{0}_{K\times KN}\,], \quad X \underline{z} = \underline{x}.
\end{gather*}

$\Phi_3$, the operator which replicates the capacitor selection into each assignment block:
\begin{equation}
  \Phi_3=\begin{bmatrix}\mathbf{0}_{K\times K}&\mathbf{0}\\\mathbf{1}_N\otimes\mathbb{I}_K&\mathbf{0}_{KN\times KN}\end{bmatrix}
\end{equation}

$\Phi_4$, the operator which sums assignments per task:
\begin{equation}
  \Phi_4=[\,\mathbf{0}_{N\times K}\;\;\mathbb{I}_N\otimes\mathbf{1}_K^T\,]
\end{equation}

$\Phi_5$, the operator which sums tasks per capacitor:
\begin{equation}
  \Phi_5=[\,\mathbf{0}_{K\times K}\;\;\mathbf{1}_N^T\otimes\mathbb{I}_K\,]
\end{equation}

\section{Circuit Setup}
\label{sec:circuit}

In \texttt{Thimble}, the source, capacitors, and the load connect through $L=2$ switched power lines (\Cref{fig:teaser}): a charge line and a load line.
Capacitors leak $\max(\kappa C_jV,\,I_0)$ as in~\S\ref{sec:costs}, with $\kappa=\SI{0.01}{\per\second}$ and $I_0=\SI{1}{\micro\ampere}$; the planner assumes $I_0=\SI{3}{\micro\ampere}$ as a pessimistic assumption.
All capacitors share the cutoff and ceiling voltages $V_{\min}=\qty{1.65}{\V}$ and $V_{\max}=\qty{3.2}{\V}$ (\Cref{tab:workload}).
Every capacitor starts at \SI{1.3}{\volt}, above the switch turn-on voltage and below $V_{\min}$, so each run begins by charging.
The source supplies up to \qty{3.3}{\V} as the working voltage of the circuit.
The sink will draw power until the voltage reaches \qty{1.65}{\V}, simulating the typical working range for microcontrollers of \qtyrange{1.65}{3.3}{\V}.
The switches are implemented as a simplification of the ROHM RZM002P02 PFET \cite{rohm_rzm002p02_2016}, consisting of a switch with an on resistance of \qty{0.1}{\ohm}.
The simplification speeds up simulation, at the cost of omitting the PFET's switching losses.
To limit switching, a capacitor at $V_{\max}$ counts as full until it falls \SI{50}{\milli\volt} below, and the running capacitor is fed at most until it first fills during each task.

\paragraph{Power scaling.}
We base our evaluation on real world energy traces, but their energy density is scaled up by $64\times$ to match the power requirements of our real world sensor node.
Similarly, we reduced the power draw of the platform's tasks by $10\times$.
With both scalings, \texttt{Thimble}'s nodes wake on every trace: on \texttt{stairs} and \texttt{office} they run almost continuously, and on \texttt{cars} intermittently.

\end{document}